\newcommand{\be}{\begin{equation}}
\newcommand{\ee}{\end{equation}}
\newcommand{\bea}{\setlength\arraycolsep{2pt} \begin{eqnarray}}
\newcommand{\eea}{\end{eqnarray}}
\newcommand{\nn}{\nonumber}

\def\ft#1#2{{\textstyle{\frac{\scriptstyle #1}{\scriptstyle #2} } }}
\def\fft#1#2{{\frac{#1}{#2}}}

\def\0{{\sst{(0)}}}
\def\1{{\sst{(1)}}}
\def\2{{\sst{(2)}}}
\def\3{{\sst{(3)}}}
\def\4{{\sst{(4)}}}
\def\5{{\sst{(5)}}}
\def\6{{\sst{(6)}}}
\def\7{{\sst{(7)}}}
\def\8{{\sst{(8)}}}
\def\sst#1{{\scriptscriptstyle #1}}

\def\M{\mathcal{M}}

\def\cH{{{\cal H}}}

\makeatletter
\@ifundefined{@parse@version@dash}{%
\def\@parse@version#1{\@parse@version@0#1}
\def\@parse@version@#1/#2/#3#4#5\@nil{%
\@parse@version@dash#1-#2-#3#4\@nil}
\def\@parse@version@dash#1-#2-#3#4#5\@nil{%
  \if\relax#2\relax\else#1\fi#2#3#4 }
}{}
\makeatother

\documentclass[%
 reprint,
 amsmath,amssymb,
 aps,
twocolumn,
preprintnumbers
]{revtex4-2}

\usepackage{graphicx}
\usepackage{dcolumn}
\usepackage{bm}
\usepackage{color}
\usepackage{tabularx}
\usepackage{hyperref}

\usepackage[vcentermath]{youngtab}
\newcommand{\myng}[1]{\,{\tiny\yng #1}\,}
\def\myyoung#1{\,{\scriptsize \young(#1)}\,}

\def\MM{\mathcal{M}}

\def\a{\alpha}

\def\e{\varepsilon}

\def\.{{\cdot}}
\def\eone{e_1}
\def\etwo{e_2}
\def\eesmall{{\scriptstyle e_1{\cdot}e_2\,}}

\def\Tr{\mbox{Tr}}
\newcommand{\lsim}{\mathrel{\hbox{\rlap{\lower.55ex \hbox{$\sim$}} \kern-.3em \raise.4ex \hbox{$<$}}}}
\newcommand{\gsim}{\mathrel{\hbox{\rlap{\lower.55ex \hbox{$\sim$}} \kern-.3em \raise.4ex \hbox{$>$}}}}

\newcommand\SO{\mathrm{SO}}
\def\be#1\ee{\begin{align}#1\end{align}}
\newcommand\s{\sigma}
\newcommand\C{\mathbb{C}}
\renewcommand\hat[1]{\widehat{#1}}
\newcommand\cD{\mathcal{D}}
\newcommand\de{\delta}
\newcommand\np{n'}
\def\tinyvdots{{\raisebox{2.5pt}{$\cdot$}\hspace{-2.35pt}\raisebox{0.5pt}{$\cdot$}\hspace{-2.35pt}\raisebox{-1.5pt}{$\cdot$}}}
\def\tinydots{{\resizebox{1em}{!}{$\cdots$}}}
\def\aone{{\raisebox{0.8pt}{$a_{\hspace{-0.2pt}1}$}}}
\def\atwo{{\raisebox{0.8pt}{$a_{\hspace{-0.2pt}2}$}}}
\def\ak{{\raisebox{0.8pt}{$a_{\hspace{-0.2pt}k}$}}}
\def\an{{\raisebox{0.8pt}{$a_{\hspace{-0.2pt}n}$}}}

\newcommand\cM{\mathcal{M}}
\newcommand\vol{\mathop{\mathrm{vol}}}

\renewcommand{\>}{\rangle}
\newcommand\p[1]{\left(#1\right)}

\definecolor{byzantine}{rgb}{0.74, 0.2, 0.64}

\newcommand\cP{\mathcal{P}}

\begin{document}

\preprint{CALT-TH 2022-17}

\title{Graviton partial waves and causality in higher dimensions}


\author{Simon Caron-Huot}
\author{Yue-Zhou Li}
\affiliation{Department of Physics, McGill University, 3600 Rue University, Montr\'eal, H3A 2T8, QC Canada}
\author{Julio Parra-Martinez}
\author{David Simmons-Duffin}
\affiliation{Walter Burke Institute for Theoretical Physics, Caltech, Pasadena, California 91125, USA}


\date{\today}

\begin{abstract}
Do gravitational interactions respect the basic principles of relativity and quantum mechanics?  We show that any graviton $S$-matrix that satisfies these assumptions cannot significantly differ from General Relativity at low energies.
We provide sharp bounds on the size of potential corrections in terms of the mass $M$ of new higher-spin states,
in spacetime dimensions $D\geq 5$ where the $S$-matrix does not suffer from infrared ambiguities.
The key novel ingredient is the full set of SO($D{-}1$) partial waves for this process,
which we show how to efficiently compute with Young tableau manipulations.
We record new bounds on the central charges of holographic conformal theories. 
\end{abstract}

\maketitle

\section{Introduction}

Relativity and quantum mechanics lie at the heart of particle physics.
Notions such as relativistic causality (``signals cannot move faster than light'')
naturally lead to the concepts of waves, fields, and particles as force carriers \cite{Weinberg:1995mt}.
Gravity challenges this unification; for example the precise meaning of causality in a fluctuating spacetime remains unclear.
In this Letter we study a
situation where causality can be unambiguously stated, and is in principle experimentally testable.

Our setup is $2\to 2$ scattering between initially well-separated objects
in a flat Minkowski-like region of spacetime.  A notion of causality
is inherited from the flat background, and encoded in the mathematically precise axioms of scattering ($S$-matrix) theory.
It can be used to constrain gravity itself.
Consider higher-derivative corrections to Einstein's gravity at long distances:
\be
S=\int \ft{d^{D}x\sqrt{-g}}{16\pi G} (R+\ft{\alpha_2}{4} C^2 +\ft{\alpha_4}{12} C^3+\ft{\alpha_4'}{6}C'^3+\dots)\,,\label{eq:action}
\ee
where $C^2,C^3, C'^3$ are higher-curvature terms defined below.
Weinberg famously argued that any theory of a massless spin-two boson must reduce to GR at long distances \cite{Weinberg:1965rz}. This was significantly extended in \cite{Camanho:2014apa},
who argued that the parameters $\alpha_i$ must be parametrically suppressed by the mass $M$ of new higher-spin states.
In parallel, $S$-matrix dispersion relations have been used to constrain signs and sizes of certain corrections \cite{Adams:2006sv,deRham:2017avq,Bellazzini:2015cra}.

Recently, by combining these methods we showed how to bound dimensionless ratios of the form $|\alpha_iM^i|$
in any scenario where $M\ll M_{\rm pl}$, such that corrections are larger than Planck-suppressed.
However, these bounds featured the infrared logarithms
that are well known to plague massless $S$-matrices in four dimensions.

In this Letter we present rigorous bounds in higher-dimensional gravity,
where infrared issues are absent.  We overcome significant technical hurdles
regarding the partial wave decompositions of higher-dimensional amplitudes.
The resulting bounds have interesting applications to holographic conformal field theories.

\section{Four-point gravity amplitudes} \label{sec:amplitudes}

\subsection{Four-point $S$-matrices and local module}

We treat the graviton as a massless particle of spin 2.
The amplitude for $2\to2$ graviton scattering
depends on the energy-momentum $p_j^\mu$ and polarization $\e_j^\mu$ of each.
It can be written generally as a sum over
Lorentz-invariant polynomials times scalar functions:
\be
 \cM = \sum_{(i)} {\rm Poly}^{(i)}(\{p_j,\e_j\})\times \M^{(i)}(s,t)\,. \label{generic M}
\ee
We use conventions in which all momenta are outgoing and Mandelstam invariants, satisfying $s+t+u=0$, are
\be
 s=\!-(p_1{+}p_2)^2,\quad  t=\!-(p_2{+}p_3)^2,\quad  u=\!-(p_1{+}p_3)^2.
\ee
In kinematics where $p_1$, $p_2$ are incoming, $s$ and $-t$
are respectively the squares of the center-of-mass energy and momentum transfer.

The allowed polynomials in \eqref{generic M} are restricted by
the fact that graviton polarizations are transverse traceless and subject to gauge redundancies \cite{Elvang:2013cua}:
\be p_j{\cdot}p_j = p_j{\cdot}\e_j = \e_j{\cdot}\e_j=0, \quad \e_j \simeq \e_j + \# p_j\,.
\ee
Depending on the choice of spanning polynomials, the functions $\M^{(i)}(s,t)$
may develop spurious singularities which would complicate their use.
As explained in \cite{Chowdhury:2019kaq}, there exist special generators of the ``local module''
such that any amplitude that is polynomial in polarizations and momenta 
leads to $\M^{(i)}$'s that are polynomial in $s$ and $t$.
These can be simply presented
using gauge- and Lorentz- invariant building blocks:
\begin{align}
\label{generators dumb}
& H_{12} = F_{1\nu}^{\mu} F_{2\mu}^\nu\,,\quad
&& H_{123} = F_{1\nu}^{\mu} F_{2\sigma}^\nu F_{3\mu}^\sigma \,,\nn\\
&
H_{1234} = F_{1\nu}^{\mu} F_{2\sigma}^\nu F_{3\rho}^\sigma F_{4\mu}^\rho\,,\quad
&& V_1 = p_{4\mu}F_{1\nu}^{\mu} p_2^\nu\,,
\end{align}
where $F_{i\nu}^\mu = p_i^\mu \e_{i\nu}-\e_i^\mu p_{i\nu}$ is proportional to the field strength.
We define $H$'s with other indices by permutation, and $V_i$ by cyclic permutations.

In this notation, any $S$-matrix involving four photons
(thus homogeneous of degree 1 in each of the vectors $\e_j^\mu$) can
be written as a sum of seven terms, involving three basic functions \cite{Chowdhury:2019kaq}:
\def\photons{4\gamma}
\be
 \MM_{\photons}=& 
 \left[ H_{14}H_{23} \MM_{\photons}^{(1)}(s,u) + X_{1243}\MM_{\photons}^{(2)}(s,u) +{\rm cyclic}\right]
\nn\\ &+ S \MM_{\photons}^{(3)}(s,t). \label{M1111}
\ee
Here, we introduced the shorthands $X$ and $S$:
\be
X_{1234} &= H_{1234}-\tfrac14H_{12}H_{34}-\tfrac14H_{13}H_{24}-\tfrac14H_{14}H_{23}\,,
\nn\\
 S &= V_1 H_{234} + V_2 H_{341} + V_3 H_{412} + V_4 H_{123}\,.
\ee
Thanks to Bose symmetry, all basic functions $\M_{\photons}^{(i)}(a,b)$ are symmetrical in their two arguments, while the third one
is further invariant under all permutations of $s,t,u$, since $S$ is fully permutation symmetric.
The combination $X$  enjoys improved Regge behavior (discussed below).

The general four-graviton amplitude $\M$ can now be written 
using all products of the photon structures, supplemented by the element $\mathcal{G}$
equal to the determinant of all dot products between $(p_1,p_2,p_3,\e_1,\e_2,\e_3,\e_4)$.
The resulting 29 generators organize under permutations as 
two singlets, seven cyclic triplets, and 
one sextuplet \cite{Chowdhury:2019kaq}:
\def\spc{,\hspace{2mm}}
\be \label{graviton basis}
{\rm singlets\!:}\ & \mathcal{G}\M^{(1)}(s,u)\spc S^2 \M^{(10)}(s,u),\nn\\
{\rm triplets\!:}\ &H_{14}^2H_{23}^2 \M^{(2)}(s,u), H_{12}H_{13}H_{24}H_{34}\M^{(3)}(s,u),\nn\\
& H_{14}H_{23}(X_{1243}{-}X_{1234}{-}X_{1324})\M^{(4)}(s,u),\nn\\
& X_{1243}^2 \M^{(6)}(s,u)\spc X_{1234}X_{1324} \M^{(7)}(s,u),\nn \\
& H_{14}H_{23} S\M^{(8)}(s,u)\spc X_{1243}S\M^{(9)}(s,u),\nonumber\\
{\rm sextuplet\!:}\  &H_{12}H_{34}X_{1243} \M^{(5)}(s,u).
\ee
These constitute a basis in generic spacetime dimension $(D\geq 8)$;
lower dimensions are reviewed in appendix~\ref{app:lowerdimlocalmodule}.

\subsection{Regge limit and dispersive sum rules}

At low energies, the effect of quartic self-interactions in the effective theory \eqref{eq:action} is to add polynomials in Mandelstam invariants to the amplitudes $\M^{(i)}$:
this is a defining property of the local module  \footnote{
We omit terms with Riemann scalar and Ricci tensors from the action,
since they are proportional to Einstein's equation of motion hence removable order-by-order in the low-energy expansion.
The structure which multiplies $C^2$ in \eqref{eq:action} is thus equivalent to the Gauss-Bonnet coupling.
}.
We would like to use the assumption that graviton scattering remains sensible at all energies
to constrain the size of these interactions.

Our axioms are best stated using smeared amplitudes:
\be \M_\Psi(s)\equiv \int_0^M dp\Psi(p)\M(s,-p^2).
\ee
As argued in \cite{Caron-Huot:2021rmr,Caron-Huot:2021enk,Caron-Huot:2022ugt,Haring:2022cyf}, for suitable wavefunctions $\Psi$,
causality is interpreted as analyticity for $s$ large in the upper-half plane,
while unitarity further implies boundedness along any complex direction:
\be
\left|\M_\Psi(s)\right|_{s\rightarrow \infty} \leq s\times\mbox{constant}\,.\label{eq: smeared Regge}
\ee
The essential conditions on $\Psi(p)$ are: finite support in $p$
(required for analyticity of $\M_\Psi$), and normalizability at large impact parameters (ensuring boundedness).

The bound \eqref{eq: smeared Regge} is assumed for polarizations that do not grow with energy.
The behavior of the scalar functions $\M^{(i)}$ 
can be deduced from the Regge scaling of the polarization structures they multiply;
leading growth rates are recorded in table \ref{table: Regge limit}.
An important observation is that the leading terms are not all linearly independent, for example
while both $X_{1234}, X_{1324}\sim s^2$, their difference grows more slowly.
The coefficients of these structures inherit the opposite behavior.
For example, the (smeared)  photon amplitudes $\M_{\photons}^{(2)}(s,t)\pm \M_{\photons}^{(2)}(u,t)$
are bounded by constants times $s^{-1}$ and $s^0$, respectively.

\begin{table}
\begin{tabular}{ |c|c|c|c|c|c|c|c|c|c|c|c| }
\hline
$H_{12}$ & $H_{13}$ & $H_{14}$ &  $X_{1234}$ & $X_{1324}$ & $X_{1243}$ & $X_{1234}{-}X_{1324}$ & $S$ & $\mathcal{G}$\\  \hline
$s^1$ & $s^1$ & $s^0$   & $s^2$ & $s^2$ & $s^1$ & $s^1$ & $s^2$ & $s^2$ \\ 
 \hline
\end{tabular}
\caption{Behavior in the fixed-$t$ Regge limit of polarization structures,
omitting some simple permutations, i.e.\ $H_{34}{\sim}H_{12}$.}\label{table: Regge limit}
\end{table}

We say that a dispersive sum rule has Regge spin $k$ if it converges assuming that $\M/s^k\to 0$;
our axioms above state that sum rules with $k>1$ converge.
As can be seen from \eqref{graviton basis} and table \ref{table: Regge limit},
$\M\sim s^k$ implies $\M^{(3)}\sim s^{k-4}$, ensuring convergence of the following
integral at fixed $t=-p^2$ (with $u=p^2-s$):
\be
 B_k^{[1]}(p^2) = \oint_{\infty}\frac{ds}{4\pi i}\left[\ft{(s-u)\M^{(3)}(s,u)}{(-su)^{\fft{k-2}{2}}}\right] \equiv 0\ \  \mbox{($k\geq 2$ even)}.
\ee
This identity yields a Kramers-Kronig type relation between scattering at low and high energies,
by a standard contour deformation argument.
Namely, one finds a low-energy contribution at the scale $M\ll M_{\rm pl}$ which is EFT-computable by assumption,
plus a discontinuity at high energies $s\geq M^2$ (see \cite{Caron-Huot:2022ugt} for more detail). See appendix~\ref{app:lowenergyamplitudes} for the low-energy amplitudes.

A salient feature of graviton scattering is that many sum rules, like $B_2^{[1]}$ above, have no denominator:
only the poles of ${\cal M}$ contribute at low energies. 
Acting on the low-energy amplitude (see \eqref{graviton exchange}), it yields:
\be \label{B1 sum rules}
8\pi G\left[\frac{1}{2p^2} + \frac{\alpha_2^2{-}2\alpha_4}{16}p^2\right]= \int_{M^2}^\infty \frac{ds}{\pi} (s-u){\rm Im}\M^{(3)}(s,u)\,.
\ee
The dependence on $p$ is exact up to EFT-computable contributions from other light poles
(such as light Kaluza-Klein modes), which we account for in our analysis below, and Planck-suppressed loop corrections, which we neglect since $M\ll M_{\rm pl}$.
Thus \eqref{B1 sum rules} constitutes an infinite number of sum rules involving two EFT parameters $\alpha_i$.
This ``superconvergence'' phenomenon is related to the graviton's spin and gauge invariance, which led
to the energy growth of structures in \eqref{graviton basis}.
For other sum rules we construct improved combinations $B_k^{\rm imp}(p^2)$ which are designed to probe finite sets of EFT couplings.
Our complete set of sum rules is detailed in appendix \ref{app:basisofsumrules}.

\section{Construction of partial waves}
\label{sec:partialwaves}

Our assumptions about the right-hand-side of  \eqref{B1 sum rules} and similar relations are minimal:
Lorentz symmetry and unitarity with respect to the asymptotic states.
The intermediate states that can appear in a scattering process in $D=d+1$ dimensions
form representations $\rho$ under $\SO(d)$ rotations in the center-of-mass frame.
Thus, the $S$-matrix can be written as a sum over projectors onto each representation.
As far as the $2\to 2$ $S$-matrix is concerned, unitarity is simply the statement
that $|S_{\rho}|\leq 1$ for the coefficient of each projector.

The main technical complication in $D>4$ is that many intermediate
representations can appear.
Furthermore, multiple index contractions can exist for a given representation.
Listing them is equivalent to enumerating on-shell three-point vertices between two massless and one massive particle.
We introduce here an efficient method to construct structures
and projectors in arbitrary $D$.

\subsection{Partial wave expansion}

Concretely, the partial wave expansion for a $2\to 2$ graviton scattering amplitude takes the form
\be
\label{eq:partialwavedecomposition}
\cM
&= s^{\frac{4-D}{2}}\sum_\rho n_\rho^{(D)} \sum_{ij} (a_{\rho}(s))_{ji}\, \pi^{ij}_\rho,
\ee
where $\rho$ runs over finite-dimensional irreps of $\SO(d)$, and the normalization $n_\rho^{(D)}$ is in \eqref{eq:ndformula}.
For completeness, a derivation of this formula is presented in appendix~\ref{app:partialwaves}.

The partial waves $\pi^{ij}_\rho$ are functions of polarizations and momenta
that transform in the representation $\rho$ under the little group $\SO(d)$ preserving $P^\mu=p_1^\mu{+}p_2^\mu$.
We build them by gluing
vertices $v^{i,a}(n,e_1,e_2)$, where $a$ is an $\SO(d)$-index for $\rho$, $i$ labels linearly-independent vertices, and
\begin{align}
\label{eq:nande}
& n^\mu \equiv \frac{p_2^\mu-p_1^\mu}{\sqrt{(p_1-p_2)^2}},\quad
e_{i}^\mu \equiv \e_i^\mu-p_i^\mu \frac{\e_i{\cdot}P}{p_i{\cdot}P}
\end{align}
are natural vectors orthogonal to $P$.
Note that $n^2=1$, and the $e_{i}$ are gauge-invariant, null, and orthogonal to $n$:
\be  n{\cdot}e_i=e_{i}^2=0\,.
\ee
In the center of mass frame, $n$ and $e_i$ are simply the orientation and polarizations
of incoming particles.
Defining an outgoing orientation similarly, $n'^\mu\propto (p_4-p_3)^\mu$,
partial waves are defined by summing over intermediate indices:
\be
\label{eq:pidefinition}
\pi^{ij}_\rho&\equiv\big( \overline{v^i},v^j\big)\equiv \overline{v^{i,a}}(n',e_3,e_4) g_{ab} v^{j,b}(n,e_1,e_2),
\ee
where $g_{ab}$ is an $\SO(d)$-invariant metric on $\rho$, and $\overline f$ denotes Schwarz reflection $\overline f(x)=(f(x^*))^*$.

Unitarity of $S$ implies that the matrix $S_\rho(s)\equiv 1+i a_\rho(s)$ satisfies $|S_\rho(s)|\leq1$, which implies $0 \leq {\rm Im}\,a_\rho \leq 2$ (where an inequality of matrices is interpreted as positive-semidefiniteness of the difference).
We illustrate these concepts in some examples in appendix~\ref{app:partialwaves}.

\subsection{Review of orthogonal representations}

A finite-dimensional irrep of $\SO(d)$ is specified by a highest weight $\rho=(m_1,\dots,m_n)$, where $n=\lfloor d/2\rfloor$, see e.g.\ \cite{Dobrev:1977qv,Kravchuk:2017dzd}. The $m$'s are integers for bosonic representations and half-integers for fermionic representations, satisfying
\be
m_1& \geq \cdots \geq m_{n-1} \geq |m_n|. 
\ee
For tensor representations, $|m_i|$ are the row lengths of the Young diagram for $\rho$. Note that $m_n$ must be positive in odd-$d$, but can be negative in even-$d$ --- the sign of $m_n$ indicates the chirality of the representation. We omit vanishing $m$'s from the end of the list, for instance denoting a spin-$J$ traceless symmetric tensor by $(J)$.

To manipulate tensors, we represent them
as index-free polynomials in polarization vectors $w_1,\dots,w_n\in \C^d$, one for each row.
The traceless and symmetry properties of a given irrep are captured by 
taking these to be orthogonal and defined modulo gauge redundancies \cite{Costa:2016hju}:
\be
w_i^2 = w_i\.w_j = 0, \quad w_j\sim w_j + \# w_i \ \ \mbox{for\  $j>i$}.
\label{eq:gaugeredundancies}
\ee
The latter means that allowed functions of $w$ must be annihilated by $w_1{\cdot}\partial_{w_2}$, etc..
Three-point vertices are then simply $\SO(d)$-invariant polynomials $v^i(w_1,\dots,w_n;n,e_1,e_2)$
where the $w$'s play the same role for a massive particle that the $\e$'s play for gravitons. 

Polynomials satisfying the gauge condition can be easily constructed
by inscribing vectors in the boxes of a Young tableau,
where each column represents an antisymmetrized product with $w$'s.
For example, given vectors $a^\mu,\ldots, e^\mu \in \C^d$, we can define a tensor in the $(3,2)$ representation via
\be
\label{eq:tableauexample}
 \myyoung{ace,bd}\equiv [w_1{\cdot}a\ w_2{\cdot}b-(a{\leftrightarrow}b)]\ [w_1{\cdot}c\ w_2{\cdot}d-(c{\leftrightarrow}d)]\ w_1{\cdot}e\,.
\ee
Any tableau defines a valid tensor.  Tableaux are not unique, since we can permute columns.
Also, antisymmetrizing all the boxes in one column with another box (of not higher height) yields a vanishing polynomial, e.g.:
\be
 \myyoung{ac,b} +  \myyoung{ba,c} +  \myyoung{cb,a} =0\,. \label{tableau id}
\ee

\subsection{Vertices with two massless and one heavy state}

With this technology, we can straightforwardly write
all three-point vertices between two gravitons and an arbitrary massive state.
Here we focus on generic dimensions $D\geq 8$, relegating special cases in lower dimensions to appendix~\ref{app:lowerdimvertices}.
All we can write are the dot product $e_1{\cdot} e_2$
and Young tableaux in which each box contains either $n$, $e_1$ or $e_2$.
Evidently, no tableau can have more than three rows, by antisymmetry.

\begin{table*}[t]\centering
\begin{tabular}{|@{}l@{}|@{}l@{}|@{}l@{}|@{}l@{}|@{}l@{}|@{}l@{}|}
\hline
\rule{0em}{0pt}
$\begin{array}{l}
 \myyoung{\bullet\bullet}\, {\scriptstyle (e_1{\cdot}e_2)^2}\\[5pt]
 \myyoung{\eone\etwo\bullet\bullet}\,\eesmall\\[5pt]
 \myyoung{\eone\eone\etwo\etwo\bullet\bullet}
\end{array}$
&
\rule{0em}{0pt}
$\begin{array}{l}
 \myyoung{\eone n\bullet\bullet,\etwo}\,\eesmall\\[10pt]
 \myyoung{\eone \eone\etwo n\bullet\bullet,\etwo}
\end{array}$
&
\rule{0em}{0pt}
$\begin{array}{l}
 {\scriptstyle (1+S)}\,\myyoung{\eone\etwo n\bullet\bullet,n}\,\eesmall\\[10pt]
 {\scriptstyle (1+S)}\,\myyoung{\eone\etwo\eone\etwo n\bullet\bullet,n}
\end{array}$
&
\rule{0em}{0pt}
$\begin{array}{l}
 \myyoung{\eone\bullet\bullet,\etwo,n}\,\eesmall\\[15pt]
 \myyoung{\eone\eone \etwo\bullet\bullet,\etwo,n}
\end{array}$
&
\rule{-0.45em}{0pt}
\rule[-3.5em]{0pt}{7.5em}
$\begin{array}{l}
\myyoung{\eone\eone\bullet\bullet,\etwo\etwo}\\[10pt]
\myyoung{\eone\etwo\bullet\bullet,nn}\,\eesmall\\[10pt]
\myyoung{\eone\etwo\eone\etwo\bullet\bullet,nn}
\end{array}$
&
\begin{tabular}{@{}l@{}}
\rule{-0.12em}{0pt}
\rule[-1.7em]{0pt}{3em}
$\myyoung{\eone\eone\etwo\bullet\bullet,\etwo nn,n}$
\rule{0.9em}{0pt}
\\
\hline
\rule{-0.14em}{0pt}
\rule[-1em]{0pt}{3em}
$\myyoung{\eone\eone\etwo\etwo\bullet\bullet,nnnn}$
\end{tabular}
\\
\hline
\rule{0.19em}{0pt}
${\scriptstyle (1+S)}\,\myyoung{\eone\eone\etwo\bullet\bullet,\etwo n}$
\rule{0.1em}{0pt}
&
\rule{0.19em}{0pt}
$\myyoung{\eone\eone\etwo n\bullet\bullet,\etwo nn}$
&
\rule{0.19em}{0pt}
${\scriptstyle (1+S)}\,\myyoung{\eone\eone\etwo\etwo n\bullet\bullet,nnn}$
&
\rule{-0.12em}{0pt}
\rule[-1.5em]{0pt}{3.5em}
$\myyoung{\eone\eone n\bullet\bullet,\etwo\etwo,n}$
&
\rule{0.19em}{0pt}
${\scriptstyle (1+S)}\,\myyoung{\eone\eone\etwo n\bullet\bullet,\etwo n,n}$
\rule{0.1em}{0pt}
&
\rule{0.19em}{0pt}
$\myyoung{\eone\eone\bullet\bullet,\etwo\etwo,nn}$
\\
\hline
\end{tabular}
\caption{The 20 graviton-graviton-massive couplings in generic dimension ($D\geq 8$).
Cells collect structures that can be in the same representation.
$\protect\myyoung{\bullet\bullet}$ stands for an arbitrary (possibly zero) even number of $n$ boxes;
$S$ flips $n$ and swaps $e_1$ and $e_2$.
\label{tab:gg8}}
\end{table*}

As a warm-up, consider two non-identical massless scalars. Two-particle states
form traceless symmetric tensors of rank $J$, i.e.\ single-row tableaux.
The only possible $\SO(d)$-invariant vertex involving $n$ is then
\be
(n{\cdot} w_1)^J = \myyoung{n\tinydots n} \quad \mbox{($J$ boxes)}.
\ee
Denoting by $\myyoung{\bullet}$ an arbitrary (possibly zero) number of boxes containing $n$,
the most general coupling between two scalars and a heavy particle is thus simply $\myyoung{\bullet}$.

Moving on to two spin-1 particles, one must add one power of each of $e_1,e_2$.
These can appear either as $e_1{\cdot}e_2$ or inside a tableau,
giving the exhaustive list:
\be
\def\spc{\hspace{2.0mm}}
 \myyoung{\bullet} \eesmall,\spc
 \myyoung{\eone\etwo\bullet}\,,\spc
\myyoung{\eone \bullet,\etwo}\,, \spc
\myyoung{\eone\etwo\bullet,n}\,, \spc
\myyoung{\eone\bullet,\etwo,n}\,,\spc
\myyoung{\eone\etwo\bullet,nn}\,. \ \label{11 couplings}
\ee
A potential tableau $\myyoung{\etwo\eone\bullet,n}$ was removed since it
is redundant thanks to \eqref{tableau id}. Thus, there are six possible vertices.
If the two particles are identical, e.g.\ photons,
we get additional restrictions on the parity in $n$ --- for example the number of boxes in the first two structures must be even.

The analogous basis of couplings for gravitons
in generic dimension $D\geq 8$ are shown in table \ref{tab:gg8}.
This basis agrees with \cite{Chakraborty:2020rxf}.
Changes in lower dimensions are listed in appendix~\ref{app:lowerdimvertices}.

\subsection{Gluing vertices using weight-shifting operators}

To glue vertices into partial waves we need to sum over intermediate spin states.
This can be achieved efficiently using weight-shifting operators \cite{Karateev:2017jgd}.
A general weight-shifting operator $\cD^{a}$ is an $\SO(d)$-covariant differential operator that carries an index $a$ for
some finite-dimensional representation of $\SO(d)$, such that acting on a tensor in the representation $\rho$
it gives a tensor in the representation with shifted weights $\rho+\de$.
We will be particularly interested in the operator $\cD^{(h)\mu}$ that removes one box at height $h$ from a Young diagram with height $h$:
\be
\cD^{(h)\mu} : \rho=(m_1,\ldots,m_h) &\to (m_1,\ldots,m_h{-}1)\equiv\rho'.
\ee
Conceptually, $\cD^{(h)\mu}$ is a Clebsch-Gordon coefficient for $\rho'\subset \myyoung{\ }\otimes \rho$:
this ensures its existence and uniqueness up to normalization.
Explicitly, $\cD^{(h)\mu}$ is given by \footnote{This weight-shifting operator was written in a different formalism in \cite{Karateev:2018oml}. To our knowledge, the expression (\ref{todorov}) in embedding coordinates $w_i$ for general $h$ is new.}
\begin{widetext}
\be
 \mathcal{D}^{(h)\mu_0} =&
 \left(\delta^{\mu_0}_{\mu_1} - \frac{w_1^{\mu_0}}{N_1^{(h)}} \frac{\partial}{\partial w_1^{\mu_1}} \right)
 \left(\delta^{\mu_1}_{\mu_2} - \frac{w_2^{\mu_1}}{N_2^{(h)}}\frac{\partial}{\partial w_2^{\mu_2}}  \right)\cdots
\left(\delta^{\mu_{h-1}}_{\mu_{h}} - \frac{w_h^{\mu_{h-1}}}{N_h^{(h)}{-}1}\frac{\partial}{\partial w_h^{\mu_{h}}}  \right)
\frac{\partial}{\partial w_{h\mu_{h}}},
\label{todorov}
\ee
\end{widetext}
where $N_i^{(h)}=d-1+m_i+m_h-i-h$.  Notice the shift by 1 in the last parenthesis: $1/(N_h^{(h)}-1)$. The $h=1$ case of~\eqref{todorov} is the familiar Todorov/Thomas operator that acts on traceless symmetric tensors \cite{Dobrev:1975ru}.

For the definition (\ref{todorov}) to be consistent,
the following properties must hold:
\begin{itemize}
\item $\cD^{(h)\mu}$ preserves the gauge constraints: for all $i<j$,
$w_i{\cdot}\partial_{w_j} \mathcal{D}^{(h)\mu}X=0$ if $X$ satisfies the same.
\item $\cD^{(h)\mu}$ sends traces to traces. By ``traces'' we mean index contractions in
strictly gauge-invariant polynomials (\emph{not} just products $w_2{\cdot}w_3$) --- 
for example, the following expression where $\mu$ denotes a unit-vector in the $\mu$ direction:
\be
\sum_{\mu=1}^{d} \myyoung{ac,b\mu,\mu}.
\ee
\end{itemize}
These properties are nontrivial and determine $\cD^{(h)\mu}$ up to an overall constant, which can be fixed by considering traces on height-$h$ columns.
For example, consider adjacent gauge transformations
$w_i{\cdot}\partial_{w_{i+1}}$. Commuting across the $i$'th and $(i+1)$'th parentheses
one finds an unwanted term proportional to $(N^{(h)}_{i}-m_i)-(N^{(h)}_{i+1}-m_{i+1}+1)$, whose
vanishing recursively determines all $N$'s in terms of $N_h^{(h)}$ as stated below~\eqref{todorov}.

Effectively, $\cD^{(h)}$ recovers indices from index-free polynomials
and enables one to evaluate the pairing \eqref{eq:pidefinition} recursively in terms of simpler pairings, for example
\begin{align}
 \!\!\left(\myyoung{a\hfill,b\hfill,c}\cdots,\myyoung{\hfill\hfill,\hfill\hfill,\hfill}\cdots\right)
= & \frac{1}{m_3}\left(\myyoung{a\hfill,b\hfill}\cdots, c{\cdot}\mathcal{D}^{(3)} \left[
\myyoung{\hfill\hfill,\hfill\hfill,\hfill}\cdots\right]\right) \nn\\
& \hspace{-3mm}+ \textrm{2 cyclic rotations of $a,b,c$.} \label{todorov pairing}
\end{align}
Such a formula holds for any choice of a column of maximal height $h$ on the left factor,
giving $1/m_h$ times a sum with alternating sign over the boxes it contains, see~(\ref{eq:integratebyparts}).
In practice, since $\cD^{(h)}$ sends tableaux to tableaux, it can be elegantly implemented as a combinatorial
operation, as discussed in appendix \ref{app:combinatorial}.

By repeatedly applying \eqref{todorov pairing} and its generalization (\ref{eq:integratebyparts}), any pairing can be reduced to a pairing between single-row tableaux of length $m_1=J$:
\be
\label{eq:examplesinglerowpairing}
\left(\myyoung{abcn\tinydots n},\myyoung{efg\np\tinydots \np}\right).
\ee
This can be computed efficiently by taking derivatives with respect to $n$ and $n'$ of the basic scalar partial wave:
\be
\label{eq:singlerowpairing}
\left(\myyoung{n\tinydots n},\myyoung{\np\tinydots \np}\right) &= (n^{\mu_1}\cdots n^{\mu_J}-\textrm{traces})(n'_{\mu_1}\cdots n'_{\mu_J}) \nn\\
&=\frac{(d-2)_J}{2^J(\frac{d-2}{2})_J} 
\cP_J\left(n\.n'\right),
\ee
where $\cP_J(x)$ is a Gegenbauer polynomial (see \eqref{eq:gegenbauer}) and $(a)_n$ is the Pochhammer symbol.
Thus, \eqref{todorov pairing} and (\ref{eq:singlerowpairing}) allow us 
to glue the vertices from table \ref{tab:gg8} into partial wave expressions which
hold for arbitrary $J{=}m_1$, involving derivatives of $\cP_J(x)$ times
dot products between graviton polarizations $e_j$ and directions $n,n'$.
This procedure can be straightforwardly and efficiently automated on a computer.

To limit the size of final expressions, we use the
Gegenbauer  equation $(x^2-1)\partial_x^2\cP_J(x)+\ldots=0$
to remove any monomial of the form $x^a\cP^{(b)}_J(x)$ with $a,b\geq 2$.
We then insert a set of linearly independent polarizations to project onto the generators
\eqref{graviton basis} of the local module and extract $\cM^{(i)}$'s that are polynomials in $x$.
Finally, we use the Gram-Schmidt method to find orthonormal combinations of vertices according to \eqref{eq:vertexnormalization}. 
As a consistency check on our results, we verified that our partial waves are
eigenvectors of the $\SO(d)$ quadratic Casimir.

\section{Results and interpretation}

Dispersive sum rules like \eqref{B1 sum rules} express low-energy EFT parameters as sums of high-energy partial waves, times unknown positive couplings.
The ``bootstrap'' game consists in finding linear combinations such that all unknowns contribute with the same sign.
Such combinations yield rigorous inequalities that EFT parameters must satisfy if a causal and unitary UV completion exists.

To obtain optimal inequalities in a gravitational setting, we follow the numerical search strategy of \cite{Caron-Huot:2021rmr,Caron-Huot:2022ugt}.
Because of the graviton pole, it is not legitimate to expand around the forward limit;
rather our trial basis consists of the improved sum rules $B_k^{\rm imp}(p^2)$ integrated against wavepackets $\psi_i(p)$ with
$|p|\leq M$. We ask for a positive action on every state of mass $m\geq M$ and arbitrary $\SO(d)$ irrep,
as well as on light exchanges of spin $J\leq 2$ and any mass.
Full details of our implementation are given in appendix \ref{app:numerics}.

\begin{figure}
\includegraphics[width=0.99\columnwidth]{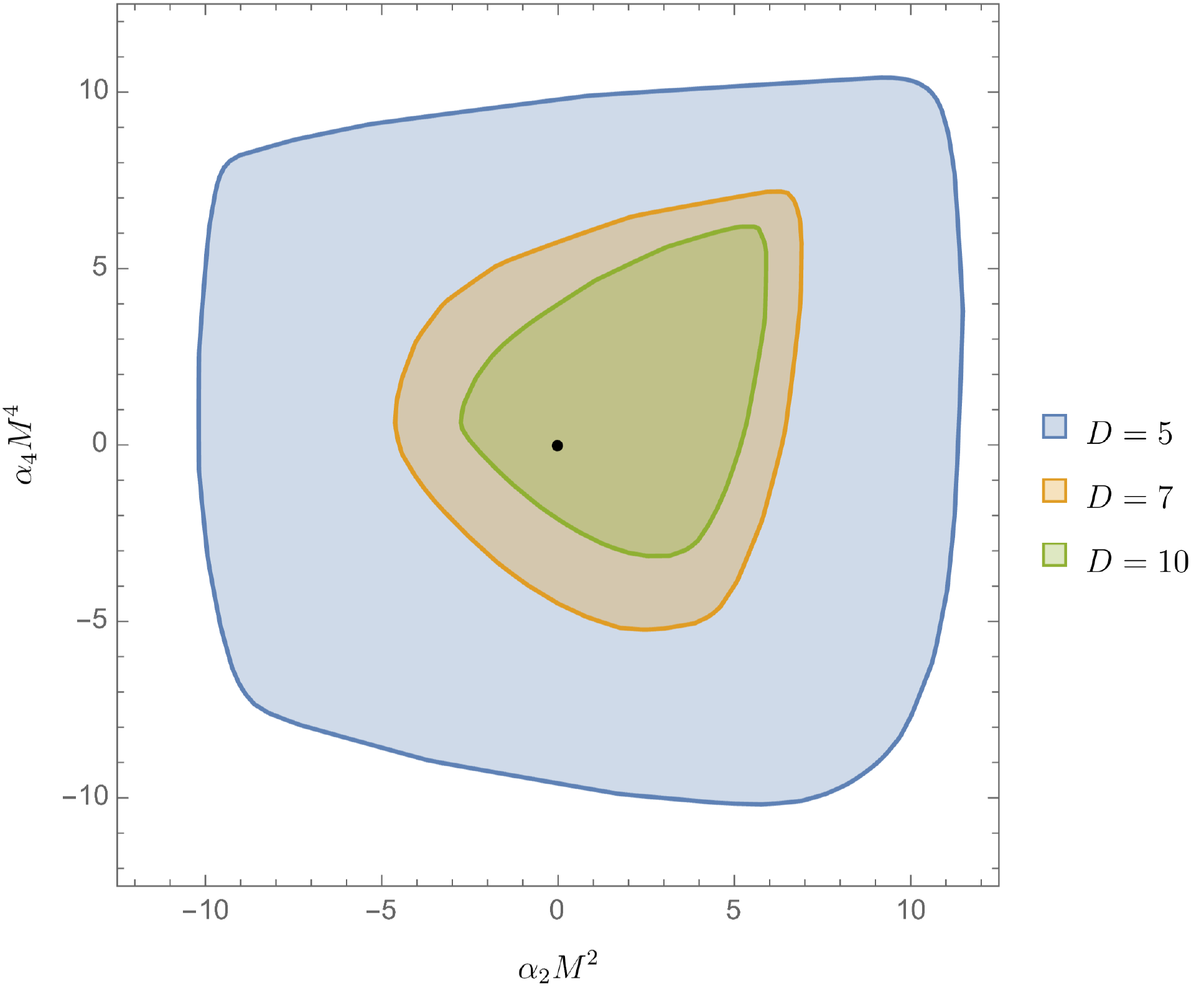} 
\caption{\label{zeplot} Allowed region for couplings $\alpha_2$ and $\alpha_4$ in $D=5$, $7$ and $10$ spacetime dimensions,
in units of the mass $M$ of higher-spin states.
}
\end{figure}

Figure \ref{zeplot} displays our main result: the allowed region for the dimensionless parameters $(\alpha_2M^2,\alpha_4M^4)$ which
control the leading corrections to the action \eqref{eq:action}, in terms of the mass $M$ of higher-spin states. For the purposes of illustration, we show the results for $D=5,7,10$; other dimensions $D$ lead to qualitatively similar plots.
The parameters are defined more precisely in \eqref{higher C},
and enter the on-shell three-graviton vertex \eqref{graviton 3pt}. It would be interesting to compare these bounds with the explicit values of Wilson coefficients in ``theory islands" arising from known UV completions \cite{BernData}.

The $M$-scaling of the bounds is significant: it implies that higher-derivative corrections can never parametrically compete with the Einstein-Hilbert term, within the regime of validity of a gravitational EFT.
As soon as corrections become significant, new particles must be around the corner. Since we assume $M\ll M_{\rm pl}$, graviton scattering is still weak at the cutoff.
In gravity, unlike in other low-energy theories, the leading (Einstein-Hilbert) interactions cannot be tuned to zero
without setting all other interactions to zero.

What happens at the scale $M$? Since we allowed for exchanges of arbitrary light states of low spins, $M$ is associated with the mass of $J\geq 3$ states. The importance of higher-spin states was anticipated in \cite{Camanho:2014apa}.
In general, higher-spin states must come in towers that include all spins \cite{Chiang:2021ziz}.
For instance, $M$ could signal the beginning of a tower of higher-spin particles (as in weakly coupled string theory),
that each couple to two gravitons with strength $\sim M^2\sqrt{G}$.
Alternatively, $M$ could be the energy at which loops representing a large number $N\sim M^{2-D}/G$ of two-particle states
that couple with weaker strength $M^{\frac{D+2}{2}}G$ to two gravitons, become non-negligible \cite{Dvali:2007hz}
\footnote{
In a Kaluza-Klein reduction from a higher dimension, $M$ can coincide with the higher-dimensional Planck mass.
Even though gravity becomes strongly interacting at that scale,
the scattering between $D$-dimensional gravitons remains weak, consistent with our bounds, since
their wavefunctions are dilute in the extra dimensions.
}.
Either way, graviton scattering must be profoundly modified at the scale $M$ and above while remaining weak.

Our flat-space bounds have implications in curved spacetimes.
As explained in \cite{Caron-Huot:2021enk}, since the scattering processes under consideration take place in a region of small size $\sim 1/M$,
flat-space dispersive bounds uplift in AdS to rigorous bounds on holographic CFTs, up to corrections suppressed by $1/(MR_{\rm AdS})=1/\Delta_{\rm gap}$.

Focusing on $D=5$ (the AdS$_5$/CFT$_4$ correspondence), stress-tensor two- and three-point functions are characterized by three parameters, including the central charges $a$ and $c$ that enter the conformal anomaly \cite{Duff:1977ay}.
Their relation to higher-derivative couplings is particularly simple when the EFT action is expressed in terms of Weyl tensors, so that renormalization of the AdS radius is avoided.
Using the field redefinition invariant formulas from \cite{Li:2021jfh} we find: 
\be
 a= \pi^2\frac{R_{\rm AdS}^3}{8\pi G},\quad \frac{a-c}{a} = \frac{2\alpha_2}{R_{\rm AdS}^2}\,.
\ee
Fig. \ref{zeplot} thus implies a sharp central charge bound:
\be
\left|\fft{a-c}{c}\right| \leq \fft{23}{\Delta_{\rm gap}^2}+\mathcal{O}(1/\Delta_{\rm gap}^4) \quad\mbox{(AdS$_5$/CFT$_4$)},
\ee
which could potentially be improved at the $\sim 5\%$ level.
In holographic theories, this result is stronger than the conformal collider bound $\frac13\leq \frac{a}{c}\leq \frac{31}{18}$ \cite{Hofman:2008ar} and establishes the parametric scaling anticipated in \cite{Heemskerk:2009pn,Camanho:2014apa,Afkhami-Jeddi:2016ntf}.
We stress that since $\Delta_{\rm gap}$ is the dimension of the lightest higher-spin (non double-trace) operator,
the bound holds even in the presence of light Kaluza-Klein modes (as in AdS$_5\times$S$_5$)
and is generally independent of the geometry of the internal manifold.
Models with $(a-c)$ of either sign are discussed in \cite{Buchel:2008vz}; our results do not exclude either sign.

The leading contact interaction in $D\geq 7$ is the 6-derivative ``third Lovelock term'', which is related to
$\alpha_4'$ in \eqref{eq:action}.
Our bounds for this coefficient depend only weakly on its sign and on $\alpha_2,\alpha_4$,
and yield the absolute limits in e.g.\ $D=7,10$:
\be
|\alpha_4' M^4|\leq 56 \ (D=7),\quad
|\alpha_4' M^4|\leq 25 \ (D=10).
\ee
In analogy with scalar EFTs \cite{Tolley:2020gtv,Caron-Huot:2020cmc,Arkani-Hamed:2020blm,Chiang:2021ziz,Albert:2022oes,Chiang:2022ltp} and four-dimensional gravitons and photons
\cite{Bern:2021ppb,Caron-Huot:2022ugt,Henriksson:2022oeu,Chiang:2022jep},
we expect this method to yield two-sided bounds on all higher-derivative interactions that can be probed by four-graviton scattering,
and on many derivative couplings involving matter fields.

\begin{acknowledgments}
We thank Cyuan-Han Chang, Clifford Cheung, Yanky Landau, Petr Kravchuk, and Sasha Zhiboedov for discussions.
DSD, SCH and YZL are supported by the Simons Foundation through the Simons Collaboration on the Nonperturbative
Bootstrap. DSD is also supported by a DOE Early Career Award under grant no.\ DE-SC0019085.
SCH is also supported by the Canada Research Chair program and the Sloan Foundation.
JPM is supported by the DOE under grant no.\ DE-SC0011632.
The computations presented here were conducted in the Resnick High Performance Computing Center, a facility supported by Resnick Sustainability Institute at the California Institute of Technology.
This research was enabled in part by support provided by Calcul Qu\'ebec and Compute Canada (Narval and Graham clusters).
\end{acknowledgments}

\bibliographystyle{apsrev4-1}
\bibliography{refs.bib} 

\onecolumngrid

\clearpage

\begin{appendix}

\section{Local module and sum rules in various dimensions}
\label{app:lowerdimlocalmodule}

\subsection{Sum rules in $D\geq 8$}
\label{app:basisofsumrules}

In $D\geq 8$, there are 19 independent
sum rules with even spin $k\geq 2$ that can be constructed from applying dispersion relations
to coefficients of the local basis with independent Regge limits \eqref{graviton basis}:
\begin{align}
 B_k(p^2)=\oint_{\infty}\fft{ds}{4\pi i}\Big\{&
\ft{(s-u)\M^{(3,10)}(s,u)}{(-su)^{\fft{k-2}{2}}},
\ft{(s-u)\M^{(2,5,8,9)+}(s,t)}{(-su)^{\fft{k-2}{2}}},
\ft{(s-u)(\M^{(6)+}(s,t)+\M^{(7)}(s,u))}{(-su)^{\fft{k-2}{2}}},
\ft{\M^{(4,6,9)-}(s,t)}{(-su)^{\fft{k-2}{2}}},
\ft{\M^{(5)-}(s,u)}{(-su)^{\fft{k-2}{2}}},
\nn\\ &
\ft{(s-u)\M^{(1,6,7,8)}(s,u)}{(-su)^{\fft{k}{2}}},
\ft{(s-u)\M^{(3)+}(s,t)}{(-su)^{\fft{k}{2}}},
\ft{(s-u)(\M^{(5)+}(t,s)-2\M^{(4)}(s,u))}{(-su)^{\fft{k}{2}}},
\ft{\M^{(5)-}(t,u)}{(-su)^{\fft{k}{2}}},
\ft{(s-u)\M^{(2)}(s,u)}{(-su)^{\fft{k+2}{2}}}
\Big\}= 0\,,
\label{eq: sum rules}
\end{align}
where $\M^{\pm} \equiv \M \pm(s\leftrightarrow u)$ and $t=-p^2=-s-u$ is held fixed. We use multiple superscripts $\cM^{(i_1,\dots,i_k)}$ to indicate a sequence of similar expressions involving the amplitudes $\cM^{(i_1)},\dots, \cM^{(i_k)}$. For odd $k>1$, there are 10 independent sum rules:
\begin{align}
 B_k(p^2)=\oint_{\infty}\fft{ds}{4\pi i}\Big\{&
\ft{\M^{(2,5,8)-}(s,t)}{(-su)^{\fft{k-3}{2}}},
\ft{\M^{(3,7)-}(s,t)}{(-su)^{\fft{k-1}{2}}},
\ft{(s-u)\M^{(4,7)+}(s,t)}{(-su){\fft{k-1}{2}}},
\ft{(s-u)\M^{(5)+}(s,u)}{(-su)^{\fft{k-1}{2}}},
\ft{(s-u)\M^{(9)}(s,u)}{(-su)^{\fft{k-1}{2}}},
 \ft{(s-u)\M^{(4)}(s,u)}{(-su)^{\fft{k+1}{2}}}
\Big\}= 0\,.
\end{align}
The Regge bound \eqref{eq: smeared Regge} implies that these sum rules converge for $k>1$.

\subsection{Sum rules in lower dimensions}
\label{eq:lowerdimsumrules}

In lower dimensions $D\leq 7$, there are two novelties for local modules as noted in \cite{Chowdhury:2019kaq}. First, new identities can reduce the number of parity-even generators of the local module. This does not occur in $D=7$. However, in $D=6$ the generator $\mathcal{G}$ does not exist, thus we must remove the parity-even sum rules involving $\M^{(1)}(s,u)$. Similarly in $D=5$, we simply remove the parity-even sum rules involving $\M^{(1,6,7)}(s,u)$.

The second novelty in lower dimensions is that new parity-odd structures appear. Following \cite{Chowdhury:2019kaq}, we organize them into multiplets under permutations. In $D=7$, there is one parity-odd singlet and two parity-odd triplets:
\be
{\rm singlets:}\ &  i S\ \epsilon(\e_1,\e_2,\e_3,\e_4,p_1,p_2,p_4)\M^{(13)}(s,u),\nn\\
{\rm triplets:}\ & i H_{14}H_{23}\,\epsilon(\e_1,\e_2,\e_3,\e_4,p_1,p_2,p_4)\M^{(11)}(s,u),\hspace{25mm} (D=7)\nn\\
& iX_{1243}\epsilon(\e_1,\e_2,\e_3,\e_4,p_1,p_2,p_4)\M^{(12)}(s,u).
\ee
Correspondingly, we can construct more sum rules
\begin{align}
& B_k(p^2)=\oint \fft{ds}{4\pi i} \Big\{\ft{\M^{(11)-}(s,t)}{(-su)^{\fft{k-2}{2}}}, \ft{(s-u)\M^{(12)}(s,u)}{(-su)^{\fft{k}{2}}}\Big\}= 0
&&(\textrm{even $k$, $D=7$}), \nn\\
& B_k(p^2) = \oint \fft{ds}{4\pi i}\Big\{\ft{\M^{(12)-}(s,t)}{(-su)^{\fft{k-1}{2}}}, \ft{(s-u)\M^{(11,12)+}(s,t)}{(-su){\fft{k-1}{2}}}, \ft{(s-u)\M^{(13)}(s,u)}{(-su)^{\fft{k-1}{2}}}\Big\}
= 0
&&(\textrm{odd $k$, $D=7$})\,.
\end{align}

In $D=6$, there are three parity-odd triplets:
\be
& H_{14}H_{23}\sigma_{1234}\cdot( V_1 \epsilon(\e_2,\e_3,\e_4,p_2,p_3,p_4))\M^{(10)}(s,u),\nn\\
& \sigma_{23}^{14}\cdot\big(\sigma_{12}^{34}\cdot (H_{24}H_{34}V_1 \epsilon(\e_1,\e_2,\e_3,p_1,p_2,p_3))\big)\M^{(11)}(s,u),\nn\\
& \sigma_{12}^{34}\cdot \big((H_{234}V_1-H_{123}V_4)\big(-p_2\cdot p_3 \epsilon(\e_1,\e_2,\e_3,\e_4,p_1,p_4)+(p_3\leftrightarrow \e_3)-(p_2\leftrightarrow \e_2)+(p_{2,3}\leftrightarrow \e_{2,3})\big)\big)\M^{(12)}(s,u).
\ee
Here, we have introduced permutation operators $\sigma$ to simplify the expressions:
\begin{align}
 \sigma_{1234}\cdot A_{1234} &\equiv A_{1234}-A_{2341}+A_{3412}-A_{4123},\nn\\
 \sigma_{ij}^{kl}\cdot A_{1234} &\equiv A_{1234} + (i\leftrightarrow j, k\leftrightarrow l)\,.
\end{align}
The corresponding parity-odd sum rules in $D=6$ are given by
\begin{align}
& B_k(p^2)=\oint \fft{ds}{4\pi i} \Big\{\ft{(s-u)\M^{(10)+}(s,t)}{(-su)^{\fft{k-2}{2}}}, \ft{\M^{(11)-}(s,u)}{(-su)^{\fft{k-2}{2}}}, \ft{\M^{(12)}(s,u)}{(-su)^{\fft{k-2}{2}}}\Big\}= 0
&&(\textrm{even $k$, $D=6$})\,, \nn\\
& B_k(p^2)=\oint_{\infty}\fft{ds}{4\pi i}\Big\{ \ft{\M^{(11)}(s,u)}{(-su)^{\fft{k-3}{2}}}, \ft{\M^{(10,12)-}(t,s)}{(-su)^{\fft{k-3}{2}}},\ft{\M^{(8)}(s,u)}{(-su)^{\fft{k-1}{2}}},,\ft{(s-u)\M^{(10,12)+}(t,s)}{(-su)^{\fft{k-1}{2}}}
\Big\}= 0
&&(\textrm{odd $k$, $D=6$})\,.
\end{align}

Finally, in $D=5$ there is one parity-odd triplet 
\be
-i \sigma_{12}^{34}\cdot\big(\sigma_{14}\cdot (H_{23}H_{234}V_1 \epsilon(\e_1,\e_4,p_1,p_2,p_4))\big)\M^{(8)}(s,u),\hspace{20mm}(D=5)
\ee
which gives rise to three independent sum rules:
\be
 B_k(p^2)=\oint_{\infty}\fft{ds}{4\pi i}\Big\{&\ft{(s-u)\M^{(8)-}(t,s)}{(-su)^{\fft{k-2}{2}}},\ft{\M^{(8)}(s,u)}{(-su)^{\fft{k-2}{2}}}
\Big\}= 0
&&(\textrm{even $k$, $D=5$})\,,\nn\\
B_k(p^2)=\oint_{\infty}\fft{ds}{4\pi i}\Big\{&\ft{\M^{(8)-}(t,s)}{(-su)^{\fft{k-3}{2}}}\Big\}= 0
&&(\textrm{odd $k$, $D=5$})\,. \label{eq:last sum rules}
\ee

\subsection{Improved sum rules}

Eqs.~\eqref{eq: sum rules}-\eqref{eq:last sum rules} provide complete sets of dispersive sum rules in the considered dimensions. By ``complete'' we mean that any sum rule with spin-$k$ convergence
can be expressed as finite sum of the $B_{\leq k}$ up to corrections that vanish faster than spin-$k$ at high energies. 
Generically, the action of $B_k(p^2)$ on the low-energy amplitude \eqref{graviton exchange} yields an infinite series of contact interactions.
Following the method in \cite{Caron-Huot:2021rmr}, all but a finite number of contacts can be removed by adding
an infinite series of higher-spin sum rules $B_{>k}^{(n)}(0)$ expanded around the forward limit.
As further discussed in \cite{Caron-Huot:2022ugt}, while it is not allowed to expand $k=2$ sum rules in the forward limit (due to the graviton pole),
there are no analogous problems for $k>2$.
Explicit formulas for the resulting $B_k^{\rm imp}(p^2)$ sum rules are recorded in ancillary files.

\section{Vertices in lower dimensions}
\label{app:lowerdimvertices}

In the main text, we described three-point vertices for two gravitons and a massive state in dimensions $D\geq 8$. In lower spacetime dimensions, the counting of three-point structures is modified, and we must take into account additional ingredients in the representation theory of the little group $\SO(d)$ (where $d=D-1$). In this section, we describe these ingredients, and then discuss the individual cases $D=7,6,5$ in turn. Detailed expressions can be found in the ancillary files included with this work.

\subsection{Representation theory ingredients}

\subsubsection{Self-duality and $\epsilon$-symbols}

When $d=2n$ is even, representations with full-height Young diagrams split into self-dual or anti-self-dual cases, according to whether $m_n$ is positive or negative. Let us explain how to account for this in our index-free formalism. Recall that the polarization vectors $w_i$ satisfy the orthogonality conditions and gauge redundancies (\ref{eq:gaugeredundancies}). When $d$ is even, the variety defined by these conditions (called a ``flag variety") splits into two irreducible components $V_\pm$, distinguished by whether $w_1\wedge\cdots \wedge w_n$ is self-dual or anti-self-dual. Specifically, we have
\be
\label{eq:conditionforvplusminus}
\frac{i^n}{n!}\epsilon_{\nu_1\cdots\nu_n}{}^{\mu_1\cdots\mu_n} w_1^{\nu_1}\cdots w_n^{\nu_n} &= \pm w_1^{[\mu_1}\cdots w_n^{\mu_n]} \quad \textrm{on $V_\pm$}.
\ee
To see why there are two components $V_\pm$, we can recursively solve the orthogonality conditions $w_i\.w_j=0$. First, we use $\SO(d)$-invariance and rescaling to set $w_1=(1,i,0,\dots,0)$. Using gauge-redundancies and $w_1\.w_i=0$, the remaining $w_i$ must have the form $w_i=(0,0,w_i^\perp)$, where $w_i^\perp\in \C^{d-2}$ are null vectors. The $w_i^\perp$ satisfy precisely the conditions and gauge redundancies for the flag variety of $\SO(d-2)$. Repeating this process for the $w_i^\perp$'s, we eventually arrive at the flag variety for $\SO(2)$, parametrized by a single null vector $w_n^{\perp\cdots\perp} \in \C^2$. Up to $\SO(2)$ transformations and rescaling, there are two possible null vectors $w_n^{\perp\cdots\perp}=(1,\pm i)$, corresponding to the two components.

The following combinations thus project the polynomial \eqref{eq:tableauexample}
associated with a tableau onto its self-dual (anti-self-dual) part:
\be
\label{eq:basicexampleselfdual}
\myyoung{\aone,\tinyvdots,\an}_\pm
&=
\myyoung{\aone,\tinyvdots,\an} \pm  i^n \epsilon(w_1,\dots,w_n,a_1,\dots,a_n).
\ee
Furthermore, the product of (\ref{eq:basicexampleselfdual}) with any polynomial in the $w_i$'s is also self-dual (anti-self-dual), since it vanishes on $V_-$ ($V_+$). In general, we define a tableau with chirality $\pm$ by adding an $\epsilon$ term to any full-height column, for example:
\be
\label{eq:youngdiagwithplusorminus}
\myyoung{adg,be,cf}_\pm &= \left(\myyoung{a,b,c}_\pm\right) \myyoung{dg,e,f}.
\ee
Note that it doesn't matter which full-height column we choose --- the resulting polynomial is the same since it agrees on both components $V_+$ and $V_-$; this can be verified explicitly with Gram determinant identities.

\subsubsection{Counting three-point structures}

Using the methods of \cite{Kravchuk:2016qvl,Chakraborty:2020rxf}, one can show that possible three-point vertices for the representation $\rho$ are classified by the following formula:
\be
\label{eq:countingformula}
\textrm{odd $D$ or $D\geq 8$} &: \begin{cases}
(\mathrm{S}^2 {\myng{(2)}}_{d-1} \otimes \rho)^{\bullet} &\textrm{if $|\rho|$ is even} \\
(\wedge^2{\myng{(2)}}_{d-1} \otimes \rho)^{\bullet} &\textrm{if $|\rho|$ is odd}
\end{cases}
\nn\\
\textrm{even $D$} &: (\mathrm{S}^2 {\myng{(2)}}_{d-1} \otimes \rho)^{\bullet}_{(-1)^{|\rho|}} \oplus (\wedge^2 {\myng{(2)}}_{d-1} \otimes \rho)^{\bullet}_{(-1)^{|\rho|+1}}.
\ee
Here, $\myng{(2)}_{d-1}$ denotes the spin-2 representation of $\SO(d-1)$. When we tensor an $\SO(d-1)$ representation with $\rho$, we implicitly dimensionally reduce $\rho$ to an $\SO(d-1)$ representation. The notation $(\lambda)^\bullet$ denotes the $\SO(d-1)$-singlet subspace of $\lambda$,
and $(\lambda)^{\bullet}_\pm$ denotes the $\SO(d-1)$ singlet subspace with parity $\pm$. Finally, $|\rho|$ is the number of boxes in the Young diagram of $\rho$. The formula (\ref{eq:countingformula}) is useful for detecting linear dependencies between Young tableau in various spacetime dimensions.

\subsubsection{Implications of $\mathsf{CRT}$}
\label{app:crt}

$\mathsf{CRT}$ symmetry relates the $\SO(d)$ representation $\rho$ to the dual reflected representation $(\rho^R)^*$. When $d\equiv 1,2,\textrm{ or }3\mod 4 $, we have simply $(\rho^R)^*=\rho$. In this case, we can choose conventions where three-point couplings for graviton-graviton-$\rho$ vertices are real,
simply by making the couplings invariant under $p_j^\mu \mapsto -p_j^\mu, i\mapsto -i$.
In particular, when computing positivity bounds, we impose that the contribution of each type of partial wave to a sum rule is a positive-definite real symmetric matrix.

Meanwhile, when $d\equiv 0\mod 4$, dual reflection changes the sign of the weight $m_n$, and hence exchanges self-dual and anti-self-dual representations $\rho_+\leftrightarrow \rho_-$. In this case, $\mathsf{CRT}$ implies that three-point coefficients of $\rho_+$ and $\rho_-$ are complex conjugates of each other. We discuss the implications of this for positivity bounds in $D=5$ below.

\subsection{Vertices in $D=7$ ($d=6$)}

Because $d=6$ is even, representations with height-3 Young diagrams split into self-dual and anti-self-dual cases.  The only effect is to double the number of height-3 tableaux in table~\ref{tab:gg8} by adding a $\pm$ chirality to each. 

Let us denote a self-dual (anti-self-dual) representation by $\rho_+$ ($\rho_-$). In the absence of parity symmetry, the three-point amplitudes $g_{gg\rho_\pm}$ between two gravitons and states in $\rho_+$ or $\rho_-$ need not be related. Consequently, we must sum over partial waves for each type of representation $\rho_+$ and $\rho_-$ independently. In bootstrap calculations, this requires including separate positivity conditions for $\rho_+$-exchange and $\rho_-$-exchange.

However, the contributions of $\rho_+$-exchange and $\rho_-$-exchange to parity-even sum rules are identical. Thus, when computing bounds using parity-even sum rules (such as our bounds on $\a_2$ and $\a_4$), positivity conditions associated to $\rho_+$ and $\rho_-$ are redundant, and it suffices to include only one of them (say $\rho_+$).

\subsection{Vertices in $D=6$ ($d=5$)}

In spacetime dimension $D=6$, $\SO(5)$ Young tableaux can have at most two rows. Since vertices are functions of five vectors $(w_1,w_2,e_1,e_2,n)$, there is a unique way to use the Levi-Civita tensor. It is convenient to write it as a height-3 column:
\be
\myyoung{\eone,\etwo,n} &\equiv \epsilon(w_1,w_2,e_1,e_2,n) \qquad \textrm{for $\SO(5)$}.
\ee
At most one column can have height 3, due to a Gram determinant identity. With this convention, the only change to table~\ref{tab:gg8} is to remove the tableau for $(J,2,2)$, and to reinterpret the tableaux
for $(J,1,1)$, $(J,2,1)$, and $(J,3,1)$ as parity-odd vertices for $(J,1)$, $(J,2)$, and $(J,3)$ respectively.

\subsection{Vertices in $D=5$ ($d=4$)}

In spacetime dimension $D=5$, $\SO(4)$ tableaux with two rows can have chirality $\pm$. In addition, we can use the Levi-Civita tensor in the form $\epsilon(w_1,a,b,c)$. Due to Gram determinant identities, this term can never be used if two-row columns are present, and it cannot be used twice. It is again convenient to draw it as a 3-row column:
\be
\label{eq:heightthreecolumnsofour}
\myyoung{\eone,\etwo,n} &\equiv \epsilon(w_1,\eone,\etwo,n) \qquad \textrm{for $\SO(4)$}.
\ee
With this convention, the tableau with row lengths $(J,1,1)$ get reinterpreted as a parity-odd coupling for the representation $\rho=(J)$. Note also that the counting formula (\ref{eq:countingformula}) implies that there are only two linearly-independent vertices for the representations $(J,\pm 2)$ with even $J$. Overall, the possible vertices in $D=5$ are given in table (\ref{tab:gg5}).

\begin{table*}[h]\centering
\begin{tabular}{|@{}l@{}|@{}l@{}|@{}l@{}|@{}l@{}|@{}l@{}|}
\hline
\rule{0pt}{0pt}
$\begin{array}{l}
 \myyoung{\bullet\bullet}\, {\scriptstyle (e_1{\cdot}e_2)^2}\\[5pt]
 \myyoung{\eone\etwo\bullet\bullet}\,\eesmall\\[5pt]
 \myyoung{\eone\eone\etwo\etwo\bullet\bullet}
\end{array}$
&
\rule{0pt}{0pt}
$\begin{array}{l}
 \myyoung{\eone n\bullet\bullet,\etwo}_\pm\,\eesmall\\[10pt]
 \myyoung{\eone \eone\etwo n\bullet\bullet,\etwo}_{\pm}
\end{array}$
&
\rule{0pt}{0pt}
$\begin{array}{l}
 {\scriptstyle (1+S)}\,\myyoung{\eone\etwo n\bullet\bullet,n}_\pm\,\eesmall\\[10pt]
 {\scriptstyle (1+S)}\,\myyoung{\eone\etwo\eone\etwo n\bullet\bullet,n}_\pm
\end{array}$
&
\rule[-3.3em]{0pt}{7.2em}
$\begin{array}{l}
 \myyoung{\eone\bullet\bullet,\etwo,n}\,\eesmall\\[13pt]
 \myyoung{\eone\eone \etwo\bullet\bullet,\etwo,n}
\end{array}$
&
\rule{0pt}{0pt}
$\begin{array}{l}
\myyoung{\eone\etwo\bullet\bullet,nn}_\pm\,\eesmall\\[10pt]
\myyoung{\eone\etwo\eone\etwo\bullet\bullet,nn}_\pm
\end{array}$
\\
\hline
\rule{-0.1em}{0pt}
\rule[-1.3em]{0pt}{3.1em}
${\scriptstyle (1+S)}\,\myyoung{\eone\eone\etwo\bullet\bullet,\etwo n}_\pm$
\rule{0.1em}{0pt}
&
\rule{0.2em}{0pt}
$\myyoung{\eone\eone\etwo n\bullet\bullet,\etwo nn}_\pm$
&
\rule{0.2em}{0pt}
${\scriptstyle (1+S)}\,\myyoung{\eone\eone\etwo\etwo n\bullet\bullet,nnn}_\pm$
&
\rule{0.2em}{0pt}
$\myyoung{\eone\eone\etwo\etwo\bullet\bullet,nnnn}_\pm$
\rule{0.1em}{0pt}
&
\\
\hline
\end{tabular}\nonumber
\caption{The graviton-graviton-massive couplings in $D=5$, as Young tableau for $\SO(4)$. We use the same notation as in table~\ref{tab:gg8}. The meaning of the height-3 column is given in (\ref{eq:heightthreecolumnsofour}).
\label{tab:gg5}}
\end{table*}

As discussed in section~\ref{app:crt}, when $d=4$, $\mathsf{CRT}$ implies that three-point coefficients of $\rho_+$ and $\rho_-$ are complex conjugates of each other. Given a pair of representations $\rho_+$, $\rho_-$ with opposite chirality, let us denote the corresponding partial waves by $\pi_+$, $\pi_-$. The $\pi_\pm$ are Hermitian matrices indexed by vertex labels $i,j$.
Exploiting fact that generators of the local module are invariant under the $Z_2\times Z_2$ symmetry which includes the interchange between initial and final states,
we can choose conventions where
\be
\pi_+ &= \pi_-^* = \pi_-^T.
\ee
By choosing generators of the local module to be invariant under $p_j\mapsto -p_j, i\mapsto-i$, 
these relations automatically hold for all the coefficients of the projector on that basis.
A contribution from $\rho_+$-exchange to the discontinuity of the amplitude takes the form
\be
\Tr(M \pi_+),
\ee
where $M=g_+ g_+^\dag$ is a Hermitian matrix built from a vector of three-point couplings $g_+$. The three-point couplings for $\rho_-$ are complex-conjugate to $g_+$ and can be grouped into the matrix $g_-g_-^\dag = g_+^* g_+^T=M^*=M^T$. Together, $\rho_+$ and $\rho_-$-exchange thus contribute
\be
\Tr( M \pi_+ ) + \Tr (M^T \pi_- ) = \Tr( M (\pi_+ + \pi_-^T) ) = 2 \Tr(M\pi_+).
\ee
So, summing the two opposite-chirality irreps simply gives a factor of 2. In parity-even sum rules, only the real-symmetric part of $M$ and $\pi$ contributes, while for parity-odd sum rules, only the imaginary part of both contributes.
Thus, when computing bounds using parity-even sum rules (as we do in this work), we can essentially pretend that the three-point couplings are real and symmetrical.
Furthermore, we need only include positivity conditions for one chirality (say $\rho_+$), since the contributions from $\rho_-$ are redundant.

\section{Details on the partial wave decomposition}
\label{app:partialwaves}

In this appendix, we derive the properly normalized partial wave decomposition \eqref{eq:partialwavedecomposition} and illustrate it for scalars and gravitons.

\subsection{Normalized partial wave expansion}

It is helpful to view the two-particle Hilbert space as a direct integral over total momentum $P=p_1+p_2$ of Hilbert spaces $\cH_P$ with fixed $P$. Because the $S$-matrix preserves momentum, it acts within each $\cH_P$.
When $P=(E,\vec 0)$, $\cH_P$ is spanned by states $|n\>$ such that $p_1=\frac E 2(1,n)$ and $p_2=\frac E 2(1,-n)$, where $n$ is a unit vector. Let us momentarily suppress the spin of the external particles, i.e.\ consider scalars.
The inner product on $\cH_P$ is a ratio of the two-particle inner product and a momentum-conserving $\de$-function:
\be
\label{eq:hpinnerproduct}
\<n'|n\> &= \frac{\<p_3|p_1\>\<p_4|p_2\>+\<p_3|p_2\>\<p_4|p_3\>}{(2\pi)^D \de^D(p_1+p_2-p_3-p_4)}
= \frac{2^d(2\pi)^{d-1}}{s^{\frac{D-4}{2}}}\p{\de(n,n') + \de(n,-n')},
\ee
where $D=d+1$ and we have used the standard single-particle inner product
\be
\<p_3|p_1\> &= 2E_1 (2\pi)^{D-1} \de^{D-1}(\vec p_1-\vec p_3).
\ee
In (\ref{eq:hpinnerproduct}), $\de(n,n')$ is a $\de$-function on the sphere $S^{d-1}$, and $s=E^2$. The inner product (\ref{eq:hpinnerproduct}) yields a corresponding completeness relation in $\cH_P$:
\be
\label{eq:completenessrelation}
1 &= \frac{s^{\frac{D-4}{2}}}{2^{d}(2\pi)^{d-1}} \frac 1 2\int_{S^{d-1}} dn |n\>\<n|,
\ee
where the factor of $\frac 1 2$ reflects Bose symmetry $|n\>=|{-}n\>$.
Using this relation it will be straightforward to correctly normalize the partial wave amplitudes.

For scalar scattering, $\cH_P$ decomposes into a direct sum of irreducible representations $\rho$ of $\SO(d)$,
where only even-spin traceless symmetric tensors $\rho=(J)$ appear, each with multiplicity one.
In the case of graviton scattering,
the states $\cH_P$ acquire extra polarization labels $|n,e_1,e_2\>$, where $e_1,e_2$ are defined by (\ref{eq:nande}),
which adds corresponding Kronecker deltas added to the above.
More general irreps $\rho$ can appear in the decomposition of $\cH_P$, and furthermore they can have nontrivial multiplicity.

For each $\rho$, we can choose basis vectors $|i,a\>$ where $a$ is an $\SO(d)$-index for $\rho$ and $i$ is a multiplicity label.
The vertices $v^{i,a}(n,e_1,e_2)$ are proportional to the overlap of $|i,a\>$ with $|n,e_1,e_2\>$:
\be
\<i,a|n,e_1,e_2\> &\equiv \p{s^{\frac{4-D}{2}}n_\rho^{(D)}}^{\frac 1 2} v^{i,a}(n,e_1,e_2),
\ee
where the constants out front have been introduced for later convenience.
We can choose the basis to be orthonormal, $\<i,a|j,b\>=\de^{ij} g^{ab}$ where $g^{ab}$ is an $\SO(d)$-invariant metric.
Projectors on $\rho$ are then 
\be
\label{eq:defofprojector}
\Pi^{ij}_\rho &\equiv |i,a\> g_{ab} \<j,b|\,, 
\ee
where $g_{ab}$ is the inverse to $g^{ab}$. As an operator on $\cH_P$, the $2\to 2$ $S$-matrix can be expanded as a sum of projectors:
\be
\label{eq:smatrixdecomposition}
S\big|_{2\to 2} &= \sum_\rho \sum_{ij} (S_{\rho}(s))_{ji} \Pi^{ij}_\rho.
\ee
Unitarity of $S$ implies that each $S_\rho(s)$ is separately a unitary matrix $S_{\rho}(s) S_\rho(s)^\dag =1$.
Taking a matrix element of $\M=-i(S-1)$ in the basis states $|n,e_1,e_2\>$,
we obtain the partial wave decomposition of the gravity amplitudes (\ref{eq:partialwavedecomposition}):
\be
\cM &= \<n',e_3^*,e_4^*|{-}i(S-1)|n,e_1,e_2\> \nn\\
&= \sum_\rho \sum_{ij} (a_\rho(s))_{ji} \<n',e_3^*,e_4^*|\Pi^{ij}_\rho|n,e_1,e_2\> \nn\\
&= s^{\frac{4-D}{2}} \sum_\rho n^{(D)}_\rho \sum_{ij} (a_\rho(s))_{ji} \pi_\rho^{ij}, \label{eq:partialwavedecompositionApp}
\ee
where $\pi_\rho^{ij}=\bar v^{i,b} g_{ba} v^{j,a}\equiv (\bar v^i,v^j)$ and $S_\rho(s)=1+ia_\rho(s)$.

From this derivation, the normalization can be fixed simply by taking the trace
of (\ref{eq:defofprojector}) and using the completeness relation (\ref{eq:completenessrelation}):
\be
\de^{ij} \dim \rho &=  \frac{n_\rho^{(D)}}{2^{d+1}(2\pi)^{d-1}} \int_{S^{d-1}} dn\, \Tr\, (\bar v^i (n), v^j(n)) = \frac{n_\rho^{(D)}\vol S^{d-1}}{2^{d+1}(2\pi)^{d-1}} \Tr\, (\bar v^i(n), v^j (n)),
\ee
where we have used rotational-invariance to perform the integral over $n$, and $\Tr$ indicates a sum over polarization states. (We detail the precise meaning of $\Tr$ for gravitons below in~(\ref{eq:vertexnormalization}).) We choose to normalize the vertices so that
\be
\Tr\, (\bar v^i(n), v^j (n)) = \de^{ij}.
\ee
The normalization coefficient $n_\rho^{(D)}$ is thus fixed to be $\dim\rho$ divided by essentially the phase space volume:
\be
\label{eq:ndformula}
n_\rho^{(D)} &= \frac{2^{d+1} (2\pi)^{d-1} \dim \rho}{\vol S^{d-1}}.
\ee
The dimension $\dim \rho$ can be computed from standard formulas, see e.g.\ \cite{Kravchuk:2017dzd,Karateev:2018oml}. For spin-$J$ traceless symmetric tensors, we have simply
\be
\dim\,(J) = \fft{(2J+d-2)\Gamma(d+J-2)}{\Gamma(d-1)\Gamma(J+1)}\,.
\ee

\subsection{Scalar scattering}

Let us determine the precise expression for $\pi_\rho$ in the case of scalar scattering.
Since each $\rho=(J)$ appears with multiplicity 1, there is a unique vertex function
\be
v(n) &= k_J \myyoung{\bullet} = k_J(n\.w_1)^J,
\ee
up to a constant $k_J$ that we determine shortly.
The partial waves are given by
\be
\pi_J(n',n) &= k_J^2(n'_{\mu_1}\cdots n'_{\mu_J}-\textrm{traces})(n^{\mu_1}\cdots n^{\mu_J}-\textrm{traces}) = k_J^2 \frac{(d-2)_J}{2^J(\frac{d-2}{2})_J} \cP_J(x),\label{eq:scalarpartialwave}
\ee
where $x=n\.n'=1+\frac{2t}{s}$, $(a)_n$ is the Pochhammer symbol, and  $\cP_J(x)$ is a Gegenbauer polynomial given by
\be
\cP_J(x) &= {}_2F_1(-J,J+d-2,\tfrac{d-1}{2},\tfrac{1-x}{2}).
\label{eq:gegenbauer}
\ee
Our normalization condition on vertices is equivalent to $\pi_J(n,n)=1$, which fixes $k_J=\p{\frac{(d-2)_J}{2^J(\frac{d-2}{2})_J}}^{-1/2}$ since $\cP_J(1)=1$.
We finally obtain $\pi_J(n',n)=\cP_J(x)$, and \eqref{eq:partialwavedecompositionApp} recovers
the familiar partial wave expansion for scalars, see e.g.\ \cite{Correia:2020xtr}.

\subsection{Graviton scattering}

In the case of graviton scattering, the orthonormality condition used in \eqref{eq:ndformula}
can be expanded as
\be
\label{eq:vertexnormalization}
\de^{ij} = \Tr\,(\bar v^i, v^j) = \sum_{e_1,e_2} \left(v^{i}(n,e_1,e_2)^*, v^{j}(n,e_1,e_2)\right),
\ee
where $\sum_{e_1,e_2}$ denotes a sum over an orthonormal basis of polarization states, and $(u,v)=u^b g_{ba} v^a$ as before. Concretely, the sum over polarizations can be performed by replacing
\be
e_1^{*\mu}e_1^{*\nu} e_1^\rho e_1^\s &\to \frac 1 2 (\hat g^{\mu\rho} \hat g^{\nu \s} + \hat g^{\nu\rho} \hat g^{\mu \s}) - \frac{1}{D-2} \hat g^{\mu\nu} \hat g^{\rho \s} \nn\\
\hat g^{\mu\nu} &\equiv \de^{\mu\nu} - n^\mu n^\nu,
\ee
where $\mu,\nu$, etc. are $\SO(d)$ indices, and making a similar replacement for $e_2$. In practice, to obtain the vertices in the ancillary files, we began with the basis of vertices in table~\ref{tab:gg8} (and the analogous bases in $D\leq 7$), and applied the Gram Schmidt procedure using the pairing (\ref{eq:vertexnormalization}).

Let us illustrate some examples of graviton partial waves for the representation $\rho=(J,1,1)$ in spacetime dimension $D\geq 8$. As shown in Table \ref{tab:gg8}, there are two linearly-independent vertices for $(J,1,1)$. An orthonormal basis with respect to the pairing \eqref{eq:vertexnormalization} is given by
\be
v_1 = \fft{i J}{\sqrt{D}(J+2)} \myyoung{\eone\bullet\bullet,\etwo,n}\eesmall\,,\quad v_2 =
\fft{i J}{J+2}\sqrt{\fft{ (J)_2 D}{(D-1)(J+D-2)_2}} \Big(\fft{1}{D}\myyoung{\eone\bullet\bullet,\etwo,n}\eesmall+
\myyoung{\eone\eone \etwo\bullet\bullet,\etwo,n}\Big)\,.
 \ee
Gluing these vertices, we can construct partial waves, which are $2$-by-$2$ matrices indexed by the vertex labels.
For brevity, we record here only the top-left corner of this matrix $\pi^{11}_\rho$, obtained by gluing $v_1$ to itself.
We furthermore write the result in terms of contributions $\pi^{11,(i)}_\rho(s,u)$ to each of the 29 scalar amplitudes defined
in \eqref{graviton basis} through the 10 generators $\cM^{(i)}(s,u)$ and their permutations. We find that $s$-channel
exchange of $(J,1,1)$ produces
\be
\label{eq:partialwavesforJ11}
& \pi^{11,(2)}_{(J,1,1)}(t,u) = \fft{2(D-4)\cP_J'(x)}{D(J+2)(J+D-5)m^8},
&& \pi^{11,(4)}_{(J,1,1)}(u,t)=\fft{8\big((D-4)\cP_J'(x)+x \cP_J''(x)\big)}{D(J+2)(J+D-5)m^8}\,,\nn\\
& \pi^{11,(5)}_{(J,1,1)}(s,u)=\fft{8\big((D-4)\cP_J'(x)+(x+1) \cP_J''(x)\big)}{D(J+2)(J+D-5)m^8},
&& \pi^{11,(5)}_{(J,1,1)}(s,t)=\fft{8\big((D-4)\cP_J'(x)+(x-1) \cP_J''(x)\big)}{D(J+2)(J+D-5)m^8},
\ee
and all other $\pi_{(J,1,1)}^{11,(i)}$ vanish. As before, $x=1+\frac{2t}{s}$. For additional expressions for partial waves, we refer the reader to the ancillary files included with this work.

\section{Low-energy amplitudes}
\label{app:lowenergyamplitudes}

\subsection{Tree-level graviton amplitudes}

The higher-derivative interactions entering the action \eqref{eq:action} are defined as:
\be \label{higher C}
& C^2 \equiv C_{\mu\nu\rho\sigma}C^{\mu\nu\rho\sigma}\,,\quad
C^3 \equiv 3C_{\mu\nu\rho\sigma}C^{\rho\sigma}\,_{\alpha\beta}C^{\alpha\beta\mu\nu}-4C_{\mu\nu\rho\sigma}C^{\nu\alpha\sigma\beta}C_{\alpha}\,^{\mu}\,_{\beta}\,^\rho\,,\nn\\
& C'^3 \equiv
-C_{\mu\nu\rho\sigma}C^{\rho\sigma}\,_{\alpha\beta}C^{\alpha\beta\mu\nu}+2C_{\mu\nu\rho\sigma}C^{\nu\alpha\sigma\beta}C_{\alpha}\,^{\mu}\,_{\beta}\,^\rho\,.
\ee
where $C_{\mu\nu\sigma\rho}$ is the Weyl tensor (traceless part of the curvature tensor $R_{\mu\nu\sigma\rho}$).
The Weyl tensor is convenient for writing low-energy effective actions since,
as mentioned in the text, the Ricci tensor and scalar can be removed using equations of motion
and do not affect our bounds.
Thus $C^2$ is equivalent to the Gauss-Bonnet density (whose coefficient is sometimes called
$\alpha_2=\lambda_{\rm GB}$), and $C'^3$ is effectively proportional to the third Lovelock density.
The normalizations in \eqref{eq:action} have been chosen so that the on-shell three-graviton vertex agrees with \cite{Camanho:2014apa}:
\be
\cM(123) &= \sqrt{32\pi G} (\mathcal{A}_1^2 + \a_2 \mathcal{A}_1 \mathcal{A}_2 + \a_4 \mathcal{A}_2^2),
\ee
where
\be
\mathcal{A}_1 &\equiv p_1\cdot\e_3\ \e_1\cdot\e_2+p_3\cdot\e_2\ \e_1\cdot\e_3+p_2\cdot\e_1 \ \e_2\cdot\e_3, \nn\\
\mathcal{A}_2 &\equiv p_1\cdot \e_3\ p_2\cdot\e_1\ p_3\cdot\e_2.
\label{graviton 3pt}
\ee

To illustrate scattering amplitudes in the local module, we now give explicit expressions
for the 10 generating amplitudes $\M^{(i)}$ entering \eqref{graviton basis} for tree-level gravity in generic dimension $D\geq 8$.
We include here higher-derivative couplings $\alpha_2,\alpha_4$ to linear order, and the unique 6-derivative interaction $\alpha_4'$ which yields a contact term:
\begin{align}
\M^{(1)}(s,u)&=8\pi G \alpha_4'+\ldots,
&
\M^{(2)}(s,u)&=\ft{8\pi G}{stu}+\ldots\,,
\nn\\
\M^{(3)}(s,u)&=\ft{8\pi G}{stu}(2-\ft{t^2\alpha_4}{2})+\ldots\,,
&
\M^{(4)}(s,u)&=\ft{8\pi G}{stu}(4-2t \alpha_2-4su\alpha_4)+\ldots\,,
\nn\\
\M^{(5)}(s,u)&=\ft{8\pi G}{stu}(8 +2\alpha_2 u)+\ldots\,, 
&
\M^{(6)}(s,u)&=\ft{8\pi G}{stu}(4 -4\alpha_2t)+\ldots\,,
\nonumber
\\
\M^{(7)}(s,u)&=\ft{8\pi G}{stu}(8 +4\alpha_2 t)+\ldots\,, 
&
\M^{(8)}(s,u)&=\ft{8\pi G}{stu}(-2\alpha_2)+\ldots\,, 
\nn\\
\M^{(9)}(s,u)&=\ft{8\pi G}{stu}(-4\alpha_2+8\alpha_4t)+\ldots\,, 
&
\M^{(10)}(s,u)&=\ft{8\pi G}{stu}(4\alpha_4)+\ldots\,.
\label{graviton exchange}
\end{align}
All omitted terms are either quadratic in the $\alpha_2,\alpha_4$ or involve higher derivative contacts, which are simply
polynomials in Mandelstam invariants subject to the symmetries of the corresponding $\M^{(i)}$.
Complete expressions, including for lower dimensions, are recorded in ancillary files.

\subsection{Kaluza-Klein and other light exchanges}

In our bounds, we allow for tree-level exchanges of massive particles that are part of the low-energy EFT --- i.e.\ whose masses are below the cutoff scale $M$.
We refer to such particles as light; they could arise, for example, from Kaluza-Klein reduction. However, we do not actually assume anything about the existence of extra dimensions.
We do however, make a choice about which types of light states to consider, and we include all representations with $J=m_1\leq 2$.
These include symmetric tensors with spin $\leq 2$, and $k$-forms of any degree, which are the possible massless string modes in string theory.
It would be interesting to consider other possible EFT matter content; we leave this problem for future work.

Given the partial waves, it is straightforward to determine the amplitudes for light exchanges.
We look for meromorphic functions $\cM^{(i)}(s,u)$ with the appropriate symmetry properties under crossing,
and possessing simple poles in Mandelstam variables whose residues match the partial waves. As an example, consider the possible KK-mode representation $\rho=(1,1,1)$ (a $3$-form).
The partial waves expressions \eqref{eq:partialwavesforJ11} predict that only the following amplitudes have $s$-channel poles: 
\be
4\M_{(1,1,1)}^{(2)}(t,u)=\M_{(1,1,1)}^{(4)}(t,u)=\M_{(1,1,1)}^{(5)}(s,u)=\M_{(1,1,1)}^{(5)}(s,t) = \fft{8}{3Dm^8(m^2-s)} + \mbox{no $s$-poles}.
\ee
We then fill in the $t$- and $u$-channel poles using symmetries.  Since $\M^{(2,4)}$ are symmetric in their two arguments, and $\M^{(5)}$ has no symmetry,
there is in fact nothing to add. That is, 3-form exchanges in all channels are accounted for by setting the function $\M_{(1,1,1)}^{(4)}(s,u)\equiv \frac{8}{3Dm^8(m^2+s+u)}$, etc.

The light amplitudes constructed via this procedure naturally have polynomial ambiguities, which represent four-point contact interactions.
Following \cite{Caron-Huot:2022ugt}, we fix these ambiguities by demanding that light states contribute to sum rules with the minimal possible spin $k$.
The contribution of light exchanges to various sum rules is then obtained by performing the appropriate contour integrals (e.g.\ \eqref{eq: sum rules}) on these amplitudes.
Our full expressions for light exchange amplitudes, and their contributions to various sum rules, can be found in the ancillary files.

When computing bounds, we demand that the contribution of each possible light exchange is sign-definite,
so that the resulting bounds are true independently of the light content of the EFT.

\section{Details of numerical implementation and ancillary files}\label{app:numerics}

Figure \ref{zeplot} was produced by numerically searching for
combinations of the $B_k^{\rm imp}(p^2)$ sum rules whose action on every unknown state is positive,
following the strategy detailed in \cite{Caron-Huot:2022ugt}. 
The sum rules are integrated against wavepackets that are polynomials in $p$ over $p\in [0,M]$,
where we typically use 5 or 6 different exponents of $p$ for each sum rules and
reach up to Regge spin $k=5$ or $k=7$.
Our search space thus contains between 200 and 400 trial sum rules.

To test positivity, we sample the action of these sum rules on a large number of
heavy states with $m\geq M$ (and light states with $J\leq 2$),
which are distributed in spin up to $J=400$.
We typically sample their action on between 10000 and 200000 states
that have spin up to $J=400$.
We also include constraints from the $m\to\infty$ scaling limit with various $b=\frac{2J}{m}$.
For the $k=2$ sum rules, which dominate at $m\to\infty$, it is important 
that the wavepackets include an overall factor $p^\alpha (M-p)$ so the sum rules
decay at large impact parameters (like $\sim 1/b^3$ in $D=5$).
We use the SDPB solver \cite{Simmons-Duffin:2015qma,Landry:2019qug} to search for  linear combinations of the trial sum rules
which are positive on all states and establish optimal bounds on the radial distance
from the origin along various rays in the $(\alpha_2,\alpha_4)$ plane.
Since the functionals depend quadratically on the $\alpha_j$, we converge toward the boundary
by optimizing a sequence of linearized quantities.

In practice, we fix the set of functionals and increase the number of states until the bounds do not change,
keeping only those sets of functionals for which such convergence could be achieved.
In going from 5 to 6 exponents, the bounds improved by no more than a few percent.
We thus expect that the recorded bounds are conservatively correct, and likely within 5\% of being optimal.

We anticipate that the partial waves computed in this work will serve in many other studies.
We have thus prepared ``process files'' which contain the complete information used to bootstrap
each of the graviton scattering process studied in this letter: \texttt{GGGG5.m}, \texttt{GGGG6.m}, \texttt{GGGG7.m}, \texttt{GGGGd.m},
for $D=5,6,7$ and $D\geq 8$ respectively, as well as a file $\texttt{GGGG4.m}$, which characterizes the $D=4$ case studied in our earlier paper \cite{Caron-Huot:2022ugt}.
Each file contains:
\begin{itemize}
\item The basis \texttt{localbasis[GGGG[d]]} of polarization structures used throughout the file,
i.e.\ the $L$ elements generated from \eqref{graviton basis} where $L=29$ for $D\geq 8$,
written in terms of the $H$, $V$, $X$, $S$ and $\mathcal{G}$ structures defined in section \ref{sec:amplitudes}
(the latter two are denoted \texttt{HS} and \texttt{HGram} in the files).
\item A list \texttt{vertices[GG[d]]} of three-point couplings $v_i$ between two gravitons and a massive state,
written in the Young Tableau notation of sections \ref{sec:partialwaves} and \ref{app:lowerdimvertices} and
divided by the scalar factor $k_J$ of \eqref{eq:scalarpartialwave} (and $e_i$ denoted \texttt{ep[i]}).
\item On-shell three-graviton vertices \texttt{amplow[GGG[d]]}, which define higher-derivative corrections like $\alpha_2, \alpha_4$.
\item Low-energy amplitudes \texttt{amplow[GGGG[d]]}, which including tree-level graviton exchanges keeping the $\alpha_k$,
as well as contact interactions \texttt{g[p,...]} that contribute up to relatively high power $p$ in Mandelstam invariants. The coefficient $8\pi G\alpha_4'$ in the main text is given by $\texttt{g[3,0,\{GGGG[d],1\}]}$ in the process files.

\item Partial waves \texttt{partialwaves[{GG[d], GG[d]}]} which list, for each possible $\SO(d)$ irrep,
an entry \texttt{exchange[irrep,\{amplitude,channel,x\},normalizations,matrix]} with typically \texttt{channel}=$s$ and \texttt{x}$=1+\frac{2t}{s}$.
If an irrep allows $n$ independent vertices, \texttt{normalizations} is an $n\times n$ matrix and \texttt{matrix} is $n\times n\times L$,
such that their entry-wise product express the projector $\pi^{ij}$ in \texttt{localbasis[amplitude]}.
The the $a$'th derivative $\cP^{(a)}_J(x)$ with respect to $x$ of the Gegenbauer polynomial \eqref{eq:gegenbauer} is denoted as \texttt{pj[J,x,D,a]}.
Irreps are denoted from the row lengths of the Young Tableau with a formal integer $m\geq 0$;
for example $\{2m+3,1\}$ denotes the family of representations $(J,1)$ where $J\geq 3$ is odd.
Non-generic irreps with low spin, for which some vertex structures disappear and the matrix becomes smaller, are explicitly separated.
\item Light exchanges \texttt{ampKK[{GG[d],GG[d]}]}, similarly written as lists of \texttt{exchange[irrep,matrix]} for each irrep,
where the $n\times n\times L$  \texttt{matrix} gives explicit functions of Mandelstam invariants.
\item Improved sum rules \texttt{sumrules[bkimp[GGGG[d],k]]}, which give $B_k^{\rm imp}$ derived from \eqref{eq: sum rules},
in terms of amplitude labels \texttt{M[...][s,-t]} entering \texttt{localbasis[GGGG[d]]},
with arguments $\texttt[s,-t]$ that indicate which Mandelstam invariants get mapped to the independent variables $m^2,p^2$
(sum rules are then $m^2$ integrals at fixed $p^2$).
\item The actions \texttt{sumruleslow[bkimp[GGGG[d],k]]} and \texttt{sumrulesKK[bkimp[GGGG[d],k]]} of sum rules on
the \texttt{amplow} and \texttt{ampKK} low-energy data.
\end{itemize}
This constitutes the full information from which the bootstrap problem can be implemented in an automated way.

\section{Weight-shifting as a combinatorial operation}
\label{app:combinatorial}

In general, the weight-shifting operator $\cD^{(h)\mu}$ lets one ``integrate-by-parts" inside an $\SO(d)$-invariant pairing to remove a box from the left factor and replace it with $\cD^{(h)\mu}$ acting on the right factor. Specifically, we have
\be
\label{eq:integratebyparts}
(w_1^{[\mu_1} \cdots w_h^{\mu_h]}g,f) &= \frac{1}{m_h}(w_1^{[\mu_1} \cdots w_{h-1}^{\mu_{h-1}} g, \cD^{(h)\mu_h]} f),
\ee
where the Young diagram for $f$ has height $h$. This is the generalization of (\ref{todorov pairing}) in the main text. In practice, this lets us remove a box from any of the tallest columns in a pairing of Young tableau.

Given (\ref{eq:integratebyparts}), we should look for an efficient way to apply $\cD^{(h)\mu}$ to a Young tableau.
This can be accomplished with the help of the following observation: 
\begin{itemize}
\item When acting on a polynomial defined via a tableau,
the derivative in the $i$'th parenthesis in (\ref{todorov}) acts \emph{only} on columns with height exactly $i$.
\end{itemize}
This leads to a simple formula for applying $\cD^{(h)\mu}$ to a Young tableau. To state it, we first define some simple operations on columns of height $k$:
\be
S[k]^{\mu\nu} \myyoung{\aone,\atwo,\tinyvdots,\ak} &= a_{1}^{\nu} \myyoung{\mu,\atwo,\tinyvdots,\ak} + a_{2}^{\nu} \myyoung{\aone,\mu,\tinyvdots,\ak} + \dots + a_{k}^{\nu} \myyoung{\aone,\atwo,\tinyvdots,\mu},
\nn\\
T[k]^\mu  \myyoung{\aone,\atwo,\tinyvdots,\ak} &= (-1)^{k-1} a_{1}^{\mu} \myyoung{\atwo,\tinyvdots,\ak} + (-1)^{k-2} a_{2}^{\mu} \myyoung{\aone,\tinyvdots,\ak} + \dots + a_{k}^{\mu} \myyoung{\aone,\atwo,\tinyvdots}.
\ee
We define $S[k]$ and $T[k]$ to give zero when acting on columns with height $k'\neq k$. We furthermore extend them to derivations on the algebra generated by columns, so that they are linear and satisfy Leibniz rules:
\be
S[k]^{\mu\nu}(xy) &= (S[k]^{\mu\nu} x)y + x(S[k]^{\mu\nu} y), \nn\\
T[k]^\mu(xy) &= (T[k]^\mu x) y + x (T[k]^\mu y).
\ee
Finally, given a tableau $Y$, let $Y^{[k]}$ denote the product of all columns of $Y$ with height $k$. In particular, $Y$ can be decomposed as $Y=\prod_{k=1}^h Y^{[k]}$, where $h$ is the height of $Y$. We claim that the action of $\cD^{(h)}$ on $Y$ is given by
\be
\mathcal{D}^{(h)\mu_0}Y =&
 \left(\left(\delta^{\mu_0}_{\mu_1} - \frac{S[1]^{\mu_0}{}_{\mu_1}}{N_1^{(h)}}  \right)Y^{[1]}\right)
 \left(\left(\delta^{\mu_1}_{\mu_2} - \frac{S[2]^{\mu_1}{}_{\mu_2}}{N_2^{(h)}} \right)Y^{[2]}\right)\cdots
\left(\left(\delta^{\mu_{h-1}}_{\mu_{h}} - \frac{S[h]^{\mu_{h-1}}{}_{\mu_{h}}}{N_h^{(h)}{-}1}\right)T[h]^{\mu_{h}}Y^{[h]}\right).
\label{eq:applydtoy}
\ee

The virtue of (\ref{eq:applydtoy}) is that it works symbolically within the algebra generated by Young tableaux.
For example, we have
\begin{align}
& v\.\mathcal{D}^{(3)} \myyoung{ad,be,cf} =
\left(v\.T[3]-\frac{1}{N_3^{(3)}-1}v\.S[3]\.T[3]\right)\myyoung{ad,be,cf}
\nn\\
&=
v{\cdot}f\myyoung{ad,be,c} - v{\cdot}e\myyoung{ad,bf,c}+v{\cdot}d\myyoung{ae,bf,c}
+ v{\cdot}c\myyoung{da,eb,f}- v{\cdot}b\myyoung{da,ec,f} \nn\\
&+v{\cdot}a\myyoung{db,ec,f}
\frac{1}{d-4} \left[c{\cdot}f \left( \myyoung{ad,be,v}+\myyoung{da,eb,v}\right) \pm \mbox{permutations}\right]\,. \label{todorov example}
\end{align}
The first line comes from applying $v\.T[3]$ and simply sums all the ways of erasing one box,
while the second line comes from applying $v\.S[3]\.T[3]$. After including permutations, it contains 9 pairs of terms
similar to the shown pair (with $c$ replaced by $a$ or $b$, or $f$ replaced by $d$ or $e$). If we add boxes with a vector $n$ to the first row, then (\ref{eq:applydtoy}) implies
\be
\label{eq:applyingD}
 v{\cdot}\mathcal{D}^{(3)} \myyoung{adn\tinydots n,be,cf} =&
\left[v{\cdot}\mathcal{D}^{(3)} \myyoung{ad,be,cf}\right]^{\!\!\raisebox{-6pt}{\myyoung{n\tinydots n}}}
\nn\\
& -\frac{m_1-2}{d-3+m_1}
\left[n{\cdot}\mathcal{D}^{(3)} \myyoung{ad,be,cf}\right]^{\!\!\raisebox{-6pt}{\myyoung{v\tinydots n}}}\,,
\ee
where each square bracket is given by eq.~\eqref{todorov example}.

\end{appendix}

\end{document}